\documentclass[journal,letters]{IEEEtran}

\usepackage[final]{graphicx}
\graphicspath{ {./figs/} }
\usepackage[reqno]{amsmath}
\usepackage{amssymb}
\usepackage{amsthm}
\usepackage{subfig}
\usepackage{epstopdf}
\usepackage{algorithm}
\usepackage{algorithmicx}
\usepackage{algpseudocode}
\usepackage{setspace}
\usepackage[T1]{fontenc}
\usepackage{color,soul}
\usepackage{multirow}
\usepackage{array}
\usepackage{cite}
\usepackage{svg}
\usepackage{verbatim}
\usepackage{enumitem,kantlipsum}
\usepackage{blindtext}
\usepackage{makecell}

\newcommand{\head}[1]{\textbf{#1}}
\newcolumntype{C}[1]{>{\centering\arraybackslash}p{#1}}

\newcommand{\ignore}[1]{}

\begin{document}

\title{Deep Learning for Frame Error Prediction\\ using a DARPA\\ Spectrum Collaboration Challenge (SC2) Dataset\vspace{-3pt}}

\author{\large{Abu Shafin Mohammad Mahdee Jameel*, {\em Student Member, IEEE}}, \large{Ahmed P. Mohamed*, {\em Student Member, IEEE}}, \large{Xiwen Zhang*, {\em Student Member, IEEE}}, \large{Aly El Gamal, {\em Senior Member, IEEE} \vspace{-20pt}}
\thanks{The authors are with the School of Electrical and Computer Engineering of Purdue University, West Lafayette, IN (e-mail: \{amahdeej, mohame23, zhan2977, elgamala\}@purdue.edu).

*Shafin, Ahmed and Xiwen contributed equally to this work.

Part of this work was supported by DARPA and AFRL under grant no. 108818. The views, opinions, and/or findings expressed are those of the
authors and should not be interpreted as representing the official views
or policies of the Department of Defense or the U.S. Government.

Approved for Public Release, Distribution Unlimited.}}

\maketitle

\begin{abstract}
We demonstrate a first example for employing deep learning in predicting frame errors for a Collaborative Intelligent Radio Network (CIRN) using a dataset collected during participation in the final scrimmages of the DARPA SC2 challenge. Four scenarios are considered based on randomizing or fixing the strategy for bandwidth and channel allocation, and either training and testing with different links or using a pilot phase for each link to train the deep neural network. We also investigate the effect of latency constraints, and uncover interesting characteristics of the predictor over different Signal to Noise Ratio (SNR) ranges. The obtained insights open the door for implementing a deep-learning-based strategy that is scalable to large heterogeneous networks, generalizable to diverse wireless environments, and suitable for predicting frame error instances and rates within a congested shared spectrum.
\end{abstract}
\IEEEpeerreviewmaketitle

\section{Introduction}

\IEEEPARstart{W}{ireless} networks are undergoing major transformations as they face significant challenges that range from supporting an unprecedented scale of the number of users to meeting the Quality of Service (QoS) requirements of new applications like Virtual Reality (VR), intelligent transportation systems, and the Internet of Things (IoT). At the same time, computational advances - most notably in Deep Learning (DL) algorithms and supporting hardware - carry the promise of enabling new possibilities, that were not considered feasible before, such as a completely autonomous and agile wireless network that can efficiently co-exist in a complex and dynamic environment, which is affected by several other independently designed networks. Three characteristics of modern wireless communication make DL an attractive option. Firstly, modeling next generation complex and dynamic networks is difficult, and depends on assumptions which may not always hold true \cite{burchfield2009rf}. Secondly, massive amount of data can be easily collected within a small time resolution, making deep network training easier \cite{hadi2018big}. Thirdly, many of the basic operations and objectives of state of the art deep neural network architectures, like convolution (or mathematical cross-correlation) and capturing temporal correlations - as in Recurrent Neural Networks (RNN) - have been also central for the design of wireless communication systems. 

The application of DL in the wireless physical communication layer has recently witnessed intense research focus, with efforts in the fields of estimation and detection \cite{neumann2018learning}, encoding and decoding \cite{ye2020deep}, scheduling and power allocation\cite{ahmed2019deep}, modeling and identification of wireless channels \cite{wang2017spatiotemporal}, as well as modulation recognition \cite{o2016convolutional}, and interference identification \cite{zhang2019deep}. 

One major improvement in the recently introduced fifth generation standard (5G) is the possibility of spectrum sharing between 4G and 5G networks \cite{omana5g2019}. Traditional approaches to spectrum sharing are severely limited, as initial rule based sharing approaches are not receptive to dynamic adjustments \cite{raychaudhuri2003spectrum}.\ignore{Further, game theory based approaches to spectrum sharing \cite{etkin2007spectrum} often rely heavily on unrealistic assumptions.} Recently, with the move towards software defined radios, approaches incorporating machine learning based techniques for route allocation have gained attention \cite{ahmed2019deep}.

The three year DARPA Spectrum Collaboration Challenge (SC2), which concluded in 2019, was intended to facilitate the research to automate the labor-intensive inefficient process of spectrum management \cite{sc2}. Research teams competing in the challenge have heterogeneously developed intelligent radio networks that autonomously coexist with significant independence, increasing the common throughput significantly. Teams employed hand tuned expert systems, as well as machine learning algorithms to solve the problem of dynamic spectrum sharing \cite{bowyer2019reinforcement}. The data gathered by various SC2 teams have been used to investigate novel techniques, some of them incorporating DL. In \cite{nguyen2019towards}, a pre-trained DL network from the computer vision domain was demonstrated to achieve good collision detection performance. 

For environments like those emulated in SC2 that depend on collaborative spectrum utilization schemes, optimization of the transmission parameters in a software defined radio network is vital \cite{stamou2019autonomic}. State based modeling like the Markov Trace Analysis, Hidden Markov Model, and pairwise error probability model have typically been used to predict future frame errors through the state of the current and past frames, without taking into account properties of the physical channel. Channel measurement based models incorporate channel properties like path loss, fading, and signal reception \cite{barsocchi2009measurement}. Recently, in \cite{saxena2018deep}, a frame error probability prediction  algorithm in a bit-interleaved coded modulation orthogonal frequency division multiplexing (BICM-OFDM) system was proposed. Although the authors also employ a deep learning approach and achieve promising results, their experiments are based on numerical simulations. Here instead, by using the data collected in SC2, which rely on DARPA's Colosseum emulator\cite{sc2}\ignore{\cite{colosseum}}, we demonstrate the feasibility of deep learning based frame error prediction - rather than only the probability - in a more complex and dynamic scenario.

We believe that due to the following factors, the proposed deep-learning-based approach is more suitable than traditional methods for frame error prediction in a collaborative radio network operating in a dynamic heterogeneous setting:

\begin{enumerate}[leftmargin=*]
    \item Nature of the challenge: There are few common parameters between different channels, even within the same network, as spectrum usage and transmission parameters vary. Also, a single network has no control over other users or environment variables, which makes it difficult to build a statistical model with strong assumptions. 
    \item Scalability: Deep learning algorithms are optimized for big data compared to traditional approaches. As described in Sec. \ref{dataset}, the considered datasets contain millions of frames. 
    \item Generalization: As shown later in the sequel, the proposed deep learning solutions generalize to diverse environments never seen during training, which is usually one of the main concerns in a data-driven approach.
\end{enumerate}

\section{Dataset} \label{dataset}

\begin{figure}
    \centering
    \includegraphics[width=0.45\textwidth]{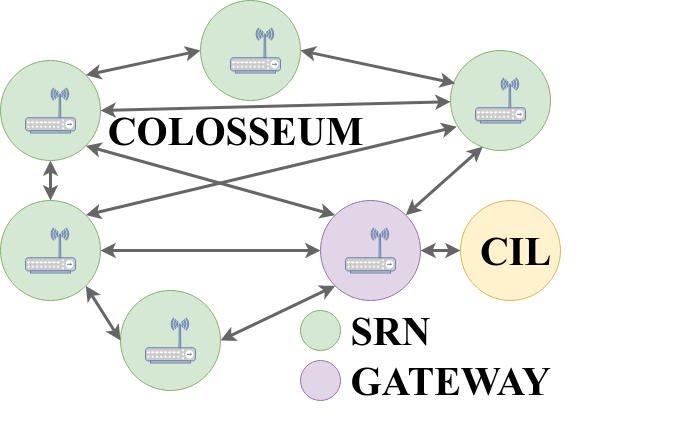}
    \caption{A DARPA SC2 Collaborative Intelligent Radio Network (CIRN) that consists of a Gateway node and multiple Software Radio Nodes (SRN). The wireless environment is emulated through DARPA's Colosseum, and different CIRNs communicate through the CIRN Interaction Language (CIL).}
    \label{fig:sysbd}
\end{figure}

The dataset is collected from the third and final Phase of SC2. A wide variety of communication environments were tested, including urban, rural, and suburban settings. A more comprehensive overview is available in \cite{sc2}.

In a match, five teams participate, and the goal is to maximize the performance of a team's Collaborative Intelligent Radio Network (CIRN) within predefined per-traffic flow QoS requirements such as maximum allowed latency and minimum allowed throughput while also keeping other teams above pre-specified thresholds. The latter enforces collaboration between teams, because no team can win simply by being greedy. Additionally, all five CIRNs can communicate over a specified data link using a common protocol, called the CIRN Interaction Language (CIL). The structure of one such CIRN is presented in Fig. \ref{fig:sysbd}. As a competitor in DARPA SC2, we constantly collected data from the 10 nodes in our CIRN\footnote{The dataset and CIRN technical design document are under review for public release. Source code for this paper is available at https://github.com/aprincemohamed/sc2-frame-error}. In total, there are 53 tables in a database file for a single match, recording all the information we collected from 10 nodes throughout a match. Here, instead of a comprehensive documentation, we only introduce the fields in this dataset that are most relevant to the considered task of frame error prediction.

\begin{itemize}[leftmargin=*]
  \item Frame Decoding Error: For every frame detected by the receiver, we record whether it has been successfully decoded, which is also the target label we try to predict.
  \item Noise Variance: For every frame, we measure and record the noise variance along the supporting radio link.
  \item Channel Allocation: We record every channel allocation update event in a match, so that for every frame transmitted, we know the allocated channel for that transmitter at the moment of transmission.
  \item Modulation and Coding Scheme (MCS): We also record the MCS selected for each frame exchanged in our CIRN. 
  \item Bandwidth (BW): We record the Bandwidth allocated to a link during transmission of a frame.
  \item Power Spectral Density (PSD): We record PSD measurements through two different sources. Every 100 ms, each node in our CIRN actively measures and records the observed PSD. In addition, peer CIRNs report their current spectrum usage, GPS information and transmit power through CIL. Based on received CIL messages, we calculate the estimated PSD for each radio node in our network.
\end{itemize}

\begin{table}[t]
\centering
\captionsetup{justification=centering}
\caption{Dataset Details for Scrimmages (Scr.) 4 and 5.}
\label{Tab:dataset}
\renewcommand{\arraystretch}{1}
\begin{tabular}{|C{1.0cm}|@{}C{1.1cm}|C{1.1cm}|C{0.8cm}|C{1.2cm}|@{}C{1.0cm}|}
	\hline
	\multirow{2}{*}{\textbf{Dataset}} & \head{No. of  Matches} &  \head{Selected Matches} & \head{Total Links} & \head{Total Frames} & \head{Frame Error} \\
	\hline
    Scr. 4 & 35  & 25 & 284 & \multicolumn{1}{|r|}{6,522,245} & \multicolumn{1}{|r|}{32.9\%}  \\
	\hline
    Scr. 5 & 100 & 59 & 627 & \multicolumn{1}{|r|}{11,669,546}  & \multicolumn{1}{|r|}{27.4\%}  \\          
	\hline
\end{tabular}
\vspace{-4mm}
\end{table}

We use two datasets collected from the last two scrimmages preceding the final event, namely Scrimmages 4 and 5. In Scrimmage 4, the factors contributing to channel allocation are randomized, e.g., relying on actually measured PSD values versus calculated values from received CIL messages. Scrimmage 5 uses a fixed allocation strategy. As a result, we note that Scrimmage 4 exhibits more frequent channel re-allocation. More specifically, we have an average of 35 channel switches per link, while the corresponding figure for Scrimmage 5 is 21 switches per link. Finally, we note that we only select matches whose scenario total bandwidth exceeds 20 MHz. Further details about the datasets are presented in Table \ref{Tab:dataset}.

\section{Deep Learning for Frame Error Prediction}

As a first example using the SC2 dataset that we collected, we carry out an empirical investigation in this section on the feasibility and potential of using deep learning for predicting frame errors. We believe that such a capability can serve as a critical component for an intelligent radio network to autonomously select the optimal transmission parameters for each radio link, e.g., MCS, transmit power and channel allocation, as well as to determine the optimal flow scheduling strategy to meet QoS requirements. 

\begin{table*}[ht!]
        \centering
        \captionsetup{justification=centering}
		\caption{Neural network architectures. For dense layers, we show the size of input and output for each layer.}
		\label{table1}
		\renewcommand{\arraystretch}{1}
		\begin{tabular}{|c|c|c|c|c|}
			\hline
			\bfseries Architecture & \bfseries \bfseries Activation Functions &\bfseries Dense Layers &\bfseries Recurrent Units\\
			\hline
            \bfseries MLP & SeLU, Softmax & \makecell {2*50 (Scr.4) or 3*50 (Scr.5), 50*50, \\ 50*50, 50*50, 50*2} &\\
            \hline
            \bfseries GRU & Sigmoid, Softmax  & 100*2 &100\\
            \hline
            \bfseries BGRU & Sigmoid, Softmax  & 200*2 &100\\
			\hline
			\bfseries BSRU & Sigmoid, Softmax  & 200*2 &100,100 (two layers)\\
			\hline
		\end{tabular}
	\end{table*}
\textbf{Problem Setup:} We naturally cast the task of frame error prediction as a binary classification problem. 
For each frame, the goal is to predict whether a frame will be decoded successfully before transmission. 
To this end, information relevant to a transmitted frame (described in Sec. \ref{dataset}), is fed to a neural network, which is employed to predict whether this frame will be decoded successfully at the receiver's side. It is important to note that all input information must be measured or collected before the frame is actually transmitted. The neural network can provide useful information to the radio network only if it can predict frame decoding errors before frames are transmitted. 

\textbf{Data Preparation:} For training, validating and testing the classifier algorithms, we employed two approaches. In the first approach, the links are divided into three sets randomly. In this case, the frames that are used for testing come from separate links compared to the training frames. The alternative method for the train-validation-test split, that we label in the figures as \emph{(with pilot)}, represents a scenario where each of the train, validation and test sets contains frames from every link. Our objective here is to quantify the impact on generalization performance when testing with links that were never seen during training. 

\textbf{Feature Selection:} In order to improve prediction performance, minimize computational cost, and alleviate overfitting, we used recursive feature elimination (RFE), which is a wrapper backward feature selection method\cite{li2017feature}. First, a decision tree classifier is trained using all features\footnote{We used the RFECV function in the Python Scikit-learn package}. Then, accuracy values obtained in absence of sets of features are used to select the feature combination that results in best classification accuracy. Both feature importance ranking and the classifier performance are stored for the final variable selection. Thereafter, the features which are the least important are removed from the present set of features. This step is recursively iterated over the present set until the best feature combination is reached. Finally, we train our neural network models using the selected features.  

\subsection{Neural Network Architectures and Training}
We started by employing a very simple deep neural network as a baseline; the fully connected multi-layer perceptron (MLP). This MLP has four hidden layers with 50 units each, and a Scaled Exponential Linear Unit (SELU) is used  as  the  activation function for the hidden layers. SELU was found to deliver better performance in this case compared to the more widely used Rectified Linear Unit (ReLU) activation function.
We then experimented with Recurrent Neural Network (RNN) architectures. First, we tested two \emph{causal} RNN networks, the first one with a single Long Short Term Memory (LSTM) layer, and the second with a single Gated Recurrent Unit (GRU) layer. We decided to use GRU as it gave better accuracy and utilizes fewer training parameters. As a result, GRU is quicker to train. Each GRU layer contains 100 GRU units, so before the output layer, a $100 \times 2$ dense layer is employed. 

In order to further improve the accuracy, we used Bidirectional RNN (BRNN) networks, which relieve the causality-in-time assumption by processing the input sequence in forward and backward directions. We note that RNN networks in general are often computationally expensive due to difficulty of parallelizing training optimization (see \cite[Chapter 10]{dl-book}). Hence, we tested two BRNN networks, one with two Simple Recurrent Unit (SRU) layers (BSRU), and the other one with a single GRU layer (BGRU), while noting that SRU units are optimized for faster training through GPU parallelization \cite{lei2018sru}. A summary of the used architectures is in Table \ref{table1}.

We have chosen softmax as the activation function in the output layer, and negative log likelihood (cross entropy) as the loss function (for a justification of these choices, see \cite[Chapter 6]{dl-book}). For MLP training, we used the stochastic gradient descent optimizer with a learning rate of .01, and for RNN-based architectures, we used Adam optimizer with a learning rate of .0001. We chose the optimizers and hyperparameters based on empirical testing. 

\subsection{Results} \label{results}
The experiments in this section are performed with a train-validation-test distribution of 40:10:50. This helps with generalizing the results compared to a train-heavy distribution. As there are more successful than erroneous frames, we also consider - in addition to the standard accuracy - the \emph{weighted accuracy}, which is defined as the average of true positive and true negative rates (sensitivity and specificity). 

We first employed the RFE feature selection method as mentioned above. The best feature combination identified for Scrimmage 4 consists of Noise Variance and MCS, while for Scrimmage 5, it contains Noise Variance, MCS and BW. We believe that BW is less relevant for Scrimmage 4 due to the randomization in channel allocation parameters. All the results presented in this section are obtained for the reduced feature set. We also tested for the full feature set, which returned reduced accuracy rates, except at low Signal to Noise Ratio (SNR) ranges for Scrimmage 5, where better accuracy was obtained.

In order to find a baseline performance with a traditional pattern recognition technique, a 3-Nearest Neighbor algorithm based on Euclidean distance to training examples was employed first, which achieved an accuracy of 66.31\% and a weighted accuracy of 61.5\% for the Scrimmage 4 dataset. 

For voice communication or streaming applications, latency becomes a critical factor. Thus, we impose a test batch size constraint of 128 frames. Note that the training batch size is 1024 frames. It is also worth mentioning that we have experimented with larger test batch sizes, which generally deliver better performance over all SNR ranges. Another interesting observation is that, when training with a very small batch size (32) and testing with a large one (1024), we obtained significantly higher true negative and significantly lower true positive rates.
\noindent
\begin{figure}[t]
\includegraphics [width=9cm]{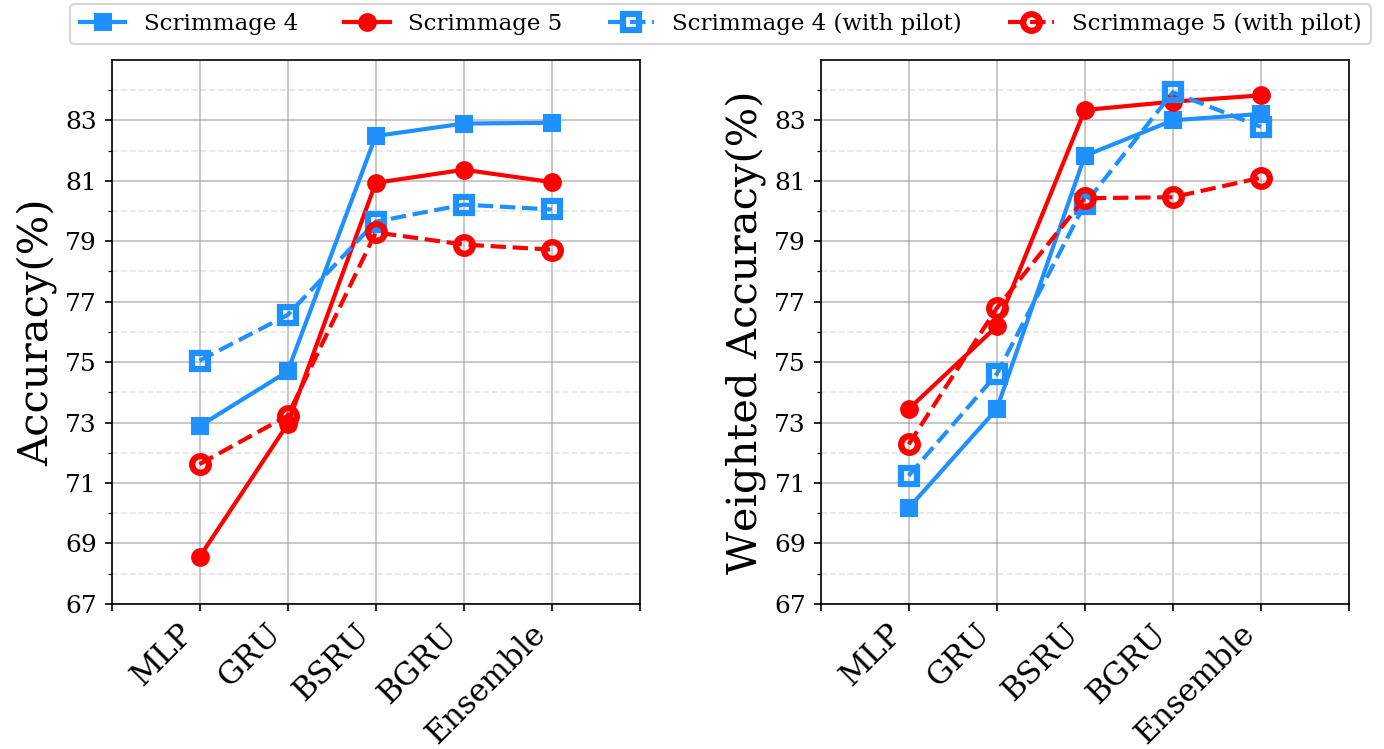}
\centering
\caption{Accuracy and weighted accuracy of different architectures for Frame Error Prediction, with and without a pilot training phase.}
\label{fig:accr}
\end{figure}
\noindent
\begin{figure}[t]
\includegraphics [width=9cm]{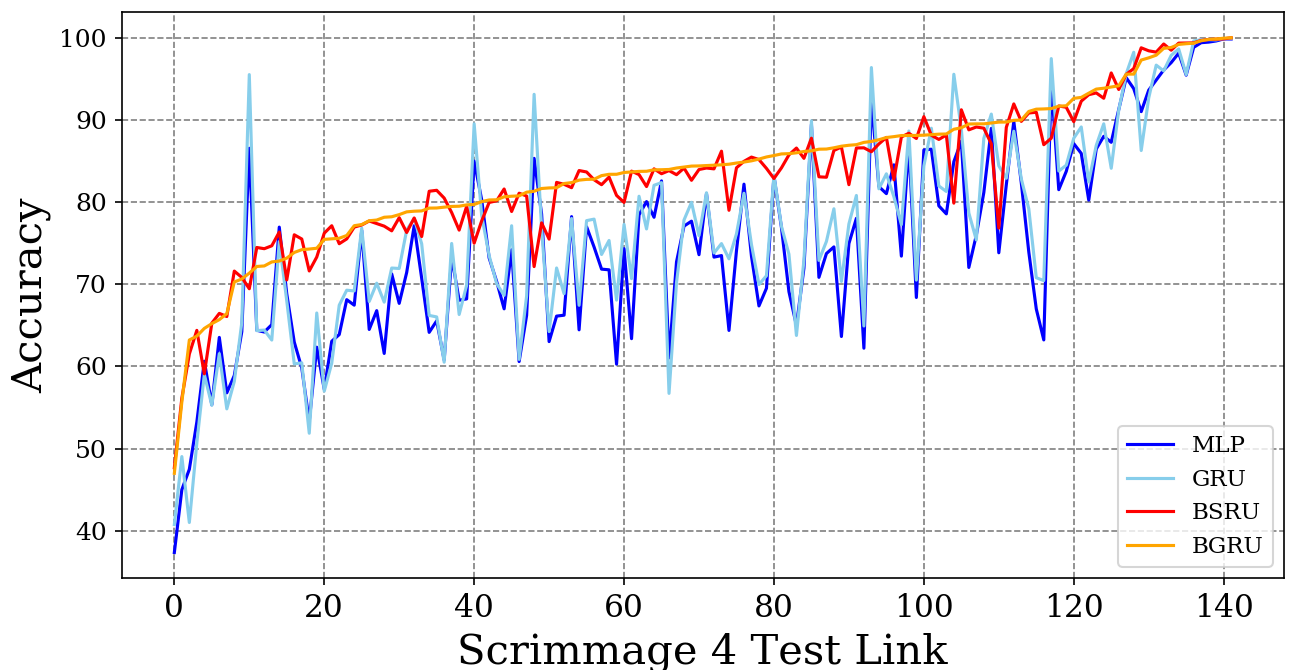}
\centering
\caption{Accuracy for individual links in Scrimmage 4.}
\label{fig:perlink}
\end{figure}
\noindent
\begin{figure}[t]
\includegraphics [width=9cm]{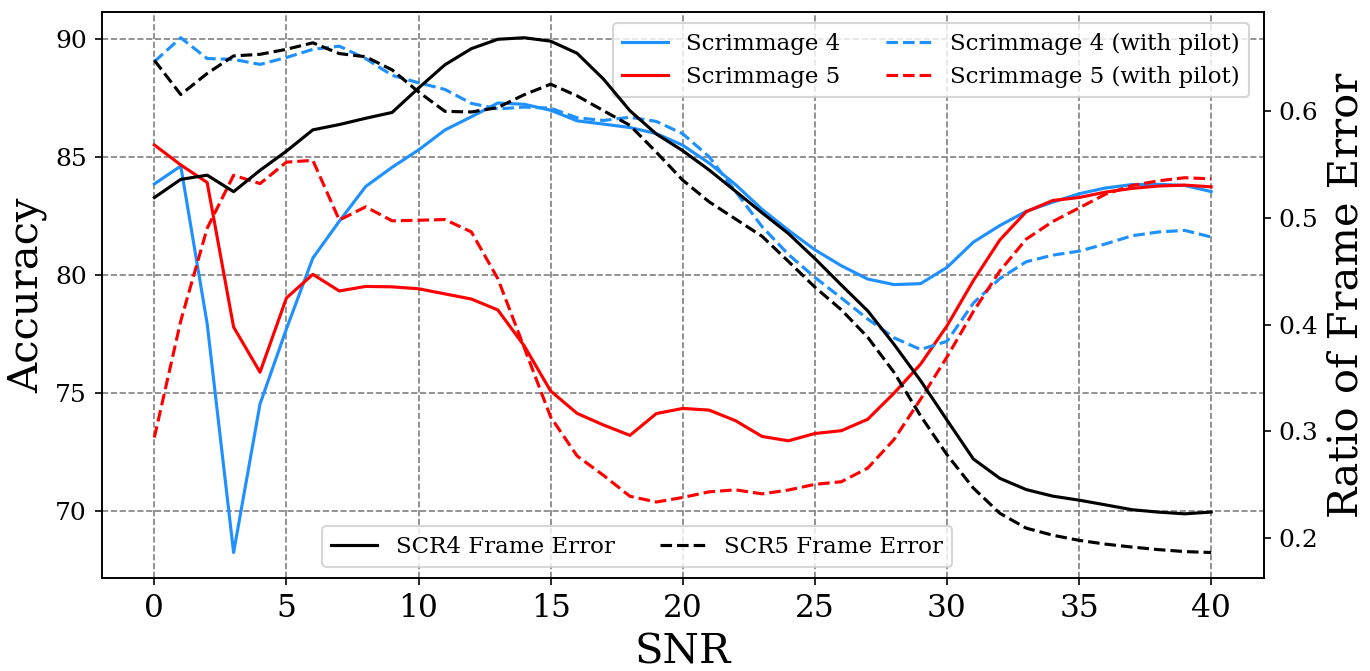}
\centering
\caption{Change of accuracy of BGRU for different SNR windows. X-axis denotes the center of a $10$ dB window. Ratio of frame errors is plotted using a separate y-axis.}
\label{fig:snr}
\end{figure}

We present the accuracy and weighted accuracy of the considered deep-learning-based frame error prediction algorithms in Fig. \ref{fig:accr}. Four experiments are performed for each deep neural network architecture, corresponding to binary choices of one of the two datasets and setting the training pilot option. In Fig. \ref{fig:perlink}, we observe the prediction performance for individual test links of Scrimmage 4. This analysis is particularly useful for a collaborative network, where link conditions may vary widely. Finally, in Fig. \ref{fig:snr}, we analyse the change in performance of a sample architecture (BGRU) over different SNR ranges. Here, the accuracy figures are reported for contiguous 10 dB SNR windows. Other architectures exhibit similar behavior. The frame error ratio over every considered SNR window is also shown on a different axis.

Based on the results in Fig. \ref{fig:accr}, Fig. \ref{fig:perlink}, Fig. \ref{fig:snr} as well as other conducted experiments, we highlight the following:
\begin{enumerate}[leftmargin=*]
    \item Randomization for Data Collection: Although randomizing the strategy for channel allocation may not be justified for real-time performance, it can unveil structures in rare events through creating diverse scenarios. We validate this hypothesis by observing that the accuracy for Scrimmage 4 (randomized allocation strategy) tend to be higher than that of Scrimmage 5 (fixed allocation strategy) for a wide SNR range (see Fig. \ref{fig:snr}). On the other hand, the Scrimmage 5 weighted accuracy is consistently higher, as observed in Fig. \ref{fig:accr}; indicating that randomization improves the sensitivity while having a negative impact on specificity. This is in spite of the higher frame error rate with randomization. However, for relatively noise free environments (SNR $>$ $32$ dB), performance gains of randomization are negligible. Finally, we noticed that the test batch size has little effect on Scrimmage 4 data, as considering a larger window of time may not be useful due to random interruptions. On the other hand, the test batch size effect on Scrimmage 5 data is more pronounced, where a bigger batch gives significantly higher accuracy at low SNR ranges.

    \item Effect of Pilot Phase: In environments with severe noise (SNR $<$ $10$ dB), training with a pilot phase is highly recommended as it provides significantly higher accuracy, specially for randomized allocation (Scrimmage 4), as seen in Fig. \ref{fig:snr}. We believe that this happens because the link conditions can vary more significantly for a randomized channel allocation strategy. As stated in Sec. \ref{dataset}, Scrimmage 4 exhibits more frequent switching of channels. Hence, adding data for the same link used for testing can significantly improve training. Further, we observe from Fig. \ref{fig:accr} that for Scrimmage 4, the pilot phase particularly improves the weighted accuracy, reflecting an enhanced ability for anomaly detection.

    \item Best Architectures:\label{item:arch} All the best performing architectures are BRNN-based, and this performance advantage is retained over all SNR ranges. We believe that this superiority of a BRNN is due to its utilization of both past and future frames in a train or test batch, providing it with current information that might only be available later due to propagation delays. This improves performance, but imposes a latency penalty. We also tested deeper versions of all architectures with more hidden layers, but that did not offer improvements in performance. Finally, from Fig. \ref{fig:perlink}, it is clear why an ensemble classifier provides very similar performance as other BRNN networks, as we observe that the two BRNN networks perform very similarly for individual links, while consistently being better than the MLP and GRU. We also tested Convolutional Neural Networks (CNN), whose performance resembles that of MLP, which is expected due to the small feature vector size. Finally, we also evaluated a Convolutional Bidirectional SRU Deep Neural Network (CBSDNN) by combining a CNN and BSRU architectures into a deep network, and the performance characteristics of that network was very close to the other BRNN networks. Results pertaining to these two architectures were omitted from the figures for brevity.
    
    \item Latency: Small test batch sizes are chosen to minimize the latency penalty imposed by BRNN. However, it is important to note that it is typical to still obtain a significant performance improvement with small test batch sizes, where it is unlikely for the propagation delay to exceed the window of time spanned by the frames in the batch.
    
    \item Training Time and Memory: Using an Nvidia Tesla P100 GPU, the total training time for 10 epochs using BGRU with Scrimmage 4 data is about 20.84 minutes, while using BSRU results in approximately a six-fold reduction in training time. However, as shown in Table \ref{tab:time}, using BSRU results in around an eight times larger GPU memory footprint. As these algorithms have comparable performance, this presents the option to optimize for either speed or memory. This can be 
    useful in future work, where we plan to implement a dynamically expandable pre-trained neural network for memory and power constrained edge devices.
    
    \item Generalization Performance: As deep neural networks are overparameterized, there is a natural risk of overfitting. Further, as our goal is to detect the relatively rare events of frame errors, this risk is increased. As we can see from Fig. \ref{fig:accr}, even when the training and testing are done with frames from separate radio links (no pilot phase) with different parameters like modulation schemes and bandwidths, the frame error prediction performance is similar or better. We also evaluated \emph{scrimmage cross-training}, where we obtained very similar results when training with Scrimmage 5 and testing on Scrimmage 4 and vice versa.
    
\end{enumerate}
\begin{table}
        \centering
        \captionsetup{justification=centering}
		\caption{Training time and GPU memory usage (Scr. 4).}
		\label{table2}
		\renewcommand{\arraystretch}{1}
		\begin{tabular}{|c|c|c|}
			\hline
			\bfseries Architecture & \bfseries \bfseries Time (Minutes) & \bfseries Memory (MiB)\\
			\hline
            \bfseries MLP & 2.03 & 675/16280\\
			\hline
            \bfseries GRU & 12.52 & 701/16280\\
            \hline
            \bfseries BGRU & 20.84 & 701/16280\\
			\hline
			\bfseries BSRU & 3.6 & 5635/16280 \\
			\hline
		\end{tabular}\label{tab:time}
	\end{table}


\section{Future Research Directions}
We presented a first example for employing deep learning for frame error prediction using our recently collected SC2 dataset. 
We next plan to explore edge artificial intelligence scenarios, where DNNs are deployed in edge communication nodes that have power and computational constraints. We hope that cloud pre-trained and dynamically expanding link-specialized neural networks would significantly improve frame error prediction performance for future collaborative spectrum sharing, where edge devices like mobile phones and AR/VR headsets rapidly acquire neural processing power.

\section*{Acknowledgment}

The authors would like to thank all members of the DARPA SC2 team, and the competitor BAM! Wireless team, for fruitful discussions and valuable help with dataset collection, with special thanks to Bharath Keshavamurthy for dataset documentation and interpretation.

\ifCLASSOPTIONcaptionsoff
  \newpage
\fi

\bibliographystyle{IEEEtran} 

\bibliography{ssp2020}

\end{document}